\documentclass[showpacs,aps, twocolumn]{revtex4-2}
\usepackage{epsfig}
\usepackage{graphicx}
\usepackage{amsmath,amssymb,amsfonts}
\usepackage{array}
\usepackage{url}
\usepackage[usenames,dvipsnames]{color}
\usepackage[colorlinks=true,linkcolor=magenta, citecolor = blue]{hyperref}
\usepackage{multirow}
\usepackage{float}
\usepackage{lineno}
\usepackage{xspace}
\usepackage{ulem}
\usepackage{lipsum} 
\usepackage{tabularx}

\begin{document}

\title{Estimating Longitudinal Polarization of $\Lambda$ and $\bar{\Lambda}$ Hyperons at Relativistic Energies using Hydrodynamic and Transport models}

\author{Bhagyarathi Sahoo}
\email{Bhagyarathi.Sahoo@cern.ch}
\author{Captain R. Singh}
\email{captainriturajsingh@gmail.com}
\author{Raghunath Sahoo}
\email{Raghunath.Sahoo@cern.ch (corresponding author)}
\affiliation{Department of Physics, Indian Institute of Technology Indore, Simrol, Indore 453552, India}

\begin{abstract}
The global and local spin polarization measurements of $\Lambda$ ($\bar{\Lambda}$) hyperons by  STAR and ALICE Collaborations open up an immense interest in investigating the spin polarization dynamics in heavy-ion collisions. Recent studies suggest the transverse component of the vorticity field is responsible for the global spin polarization. In contrast, the longitudinal component of the vorticity field accounts for the local spin polarization. The local (longitudinal) spin polarization of $\Lambda$-hyperons arises due to the anisotropic flows in the transverse plane, indicating a quadrupole pattern of the longitudinal vorticity along the beam direction. In this study, we derive a simple solution relating the longitudinal mean spin vector with the second-order anisotropic flow coefficient due to the thermal shear tensor for an ideal uncharged fluid in a longitudinal boost invariant scenario. The present study focuses on the local spin polarization of $\Lambda$ and $\bar{\Lambda}$ in Au$+$Au and Pb$+$Pb collisions at $\sqrt{s_{NN}}$ = 200 GeV and 5.02 TeV, respectively. Further, we explore the azimuthal angle, centrality, and transverse momentum ($p_{\rm T}$) dependence study of longitudinal spin polarization using hydrodynamic and transport models. All these models predict a maximum longitudinal spin polarization in mid-central collisions around 30-50 \% centrality at $p_{\rm T} \approx$  2.0 - 3.0 GeV/c. These findings on longitudinal spin polarization advocate the existence of a thermal medium in non-central heavy-ion collisions.

\end{abstract}
\date{\today}
\maketitle

\section{Introduction}
\label{intro}
Recently, many studies have been conducted to investigate the physics involved in non-central ultra-relativistic heavy-ion collisions at the Relativistic Heavy Ion Collider (RHIC) and the Large Hadron Collider (LHC). The medium created in these collisions is believed to be the hottest, most dense, least viscous, and the most vortical fluid ever found in nature~\cite{STAR:2017ckg, Bernhard:2019bmu}. The shear in the initial velocity distribution of the participants manifests a non-zero vorticity component, resulting in a net polarization of the produced particles along the orbital angular momentum (OAM) direction. This phenomenon is 
called global spin polarization~\cite{Liang:2004ph, Becattini:2015ska}. Excluding OAM contribution, other potential sources are contributing to vorticity. These include anisotropic flow in the transverse plane, jet-like fluctuation in the fireball, medium viscosity, and due to the strong magnetic field (Einstein-de Hass effect)~\cite{Xia:2018tes, Jiang:2016woz, Wei:2018zfb, Becattini:2017gcx, Pang:2016igs, Voloshin:2017kqp, Betz:2007kg, Sun:2018bjl,  Sahoo:2023xnu, Pradhan:2023rvf}. Depending on the location of fluid elements within the created system, vorticity is generated in different directions, which could induce a local spin polarization of the particles.\\

At the early stages of the development of the relativistic spin-polarization framework, most relativistic hydrodynamic models assume the spin degrees of freedom in local thermodynamical equilibrium for 
studying global spin polarization. Consequently, a relationship emerges between the spin polarization vector and the thermal vorticity, suggesting spin polarization is directly proportional to the gradient of velocity and temperature 
fields. Such spin polarization aligns with the direction of OAM. Besides global spin polarization, the local (or longitudinal) spin polarization of $\Lambda$-hyperons along the beam 
direction is observed in experiments ~\cite{STAR:2019erd, Niida:2018hfw,ALICE:2021pzu,STAR:2023eck}. The longitudinal spin polarization arises due to the inhomogeneous 
expansion of the fireball. Specifically, in non-central heavy-ion collisions, anisotropic flow in the 
transverse plane forms a quadrupole pattern of longitudinal 
vorticity along the beam direction (z-axis), leaving particles longitudinally polarized~\cite{Xia:2018tes, 
Jiang:2016woz, Wei:2018zfb, Becattini:2017gcx, Pang:2016igs, Voloshin:2017kqp}. Unlike the spin polarization along 
OAM, the longitudinal spin polarization ($P_z$) is sensitive only to transverse expansion dynamics~\cite{Becattini:2017gcx}. It does not decrease rapidly with increasing center-of-mass energy; thus, it is detectable even at LHC energies. Furthermore, it is sustained in the Bjorken longitudinal boost invariance scenario and does not require identifying the orientation of the reaction plane. However, the topology of the colliding ions plays a crucial role in determining the longitudinal polarization of the 
particles.\\

Over time, significant progress has been made in investigating the spin polarization of the particles in ultra-relativistic collisions~\cite{Liang:2004ph, Liang:2004xn, Becattini:2013vja, Becattini:2019ntv, Karpenko:2016jyx, Xie:2016fjj, Becattini:2022zvf, Hidaka:2022dmn, Sahoo:2023oid, Sahoo:2024yud, Becattini:2013fla, STAR:2017ckg, STAR:2018gyt, ALICE:2019onw, Li:2017slc, Fu:2020oxj, Alzhrani:2022dpi, Karpenko:2017lyj, Vitiuk:2019rfv, Sun:2017xhx, Fang:2016vpj, Ivanov:2019ern, Xie:2017upb, Li:2017dan, Becattini:2020ngo}. There are numerous theoretical advancements, including Liang and Wang's prediction of strange quark polarization caused due to the spin-orbit coupling within the thermalized QCD matter~\cite{Liang:2004ph}. Additionally, Becattini et al. proposed the global spin polarization of $\Lambda$-hyperons due to thermal vorticity~\cite{Becattini:2013fla}. Later, various theoretical frameworks 
based on (3+1)D relativistic hydrodynamic and multiphase transport (AMPT) models were formulated to explain the global spin polarization of $\Lambda$-hyperons. Their predictions quantitatively and qualitatively agree with the experimental findings at the RHIC~\cite{STAR:2017ckg, STAR:2018gyt, ALICE:2019onw, Li:2017slc, Fu:2020oxj, Alzhrani:2022dpi, Karpenko:2017lyj, Vitiuk:2019rfv, Sun:2017xhx, Fang:2016vpj, Ivanov:2019ern, Xie:2017upb, Li:2017dan, Becattini:2020ngo}. This convergence of theory and experiment implies a need for a thorough study of longitudinal polarization of $\Lambda$-hyperons.\\

The experimental measurement of longitudinal spin polarization of $\Lambda$ ($\bar{\Lambda}$)  with the azimuthal angle ($\phi$) relative to the second-order event plane ($\Psi_2$) by the 
STAR Collaboration~\cite{STAR:2019erd, Niida:2018hfw} shows an opposite trend compared to the theoretical predictions~\cite{Becattini:2017gcx}. This discrepancy between theoretical models and experimental data is known as the ``spin sign puzzle". A study by Xie et al. employed a (3+1)D particle-in-cell relativistic hydrodynamic model, and the obtained results qualitatively agree with the experimental data~\cite{Xie:2020}. Nonetheless, the reason for the disparate signs of longitudinal spin polarization across various hydrodynamic models remains uncertain. In some studies, the spin sign puzzle is addressed by adding a thermal shear term at local equilibrium into the spin polarization vector~\cite{Becattini:2021iol, Fu:2021pok, Liu:2021uhn, Becattini:2021suc, Palermo:2024tza}. The authors have obtained the modified spin polarization vector in the linear response theory by assuming an isothermal decoupling hypersurface in local equilibrium with linear order in the gradients of all thermodynamic fields. This modified spin polarization vector for spin-1/2 baryons contains contributions from kinematic vorticity and the kinematic shear tensor. Using this definition of the spin polarization vector in the hydrodynamic model, the authors could reasonably explain the longitudinal polarization data~\cite{Becattini:2021iol}. Further, the AMPT-based study addresses this spin sign issue by considering the quadrupole structure of longitudinal vorticity~\cite{Xia:2018tes}. \\

Moreover, there have been several additional efforts to address this issue. For instance, by substituting the thermal vorticity in the spin polarization equation (Eq.~\ref{eq1}) with projected thermal vorticity or T-vorticity, the longitudinal component of polarization flips its sign and aligns with experimental observations~\cite{Florkowski:2019voj, Wu:2019eyi}. Recently, Yi et al. estimated that the emergence of 
longitudinal spin polarization is due to a combination of shear-induced tensor, thermal vorticity, and fluid acceleration~\cite{Yi:2021ryh, Wu:2022mkr, Yi:2023tgg}. They argued that while the shear-induced term accounts for the correct angular 
dependence of longitudinal polarization, its overall impact is canceled out by thermal vorticity and fluid acceleration. Furthermore, Florkowski et al. examined the contributions of thermal 
shear and thermal vorticity to longitudinal polarization ~\cite{Florkowski:2021xvy}. In their earlier work, they 
attempted to resolve this spin sign puzzle by incorporating a spin tensor component into the orbital angular momentum~\cite{Florkowski:2017ruc, Florkowski:2017dyn, Florkowski:2018fap, Bhadury:2020cop, Bhadury:2020puc, Florkowski:2018ahw}. In their work, it is argued that if the energy-momentum tensor is not symmetric, it implies the spin tensor is not conserved. However, the total angular momentum is always conserved. As a result, the spin-polarization tensor and thermal vorticity are not equal, unlike the global equilibrium case~\cite{Florkowski:2018ahw}. Over several years, various theoretical frameworks have been developed in relativistic hydrodynamics with spin as a dynamical degree of freedom from different approaches~\cite{Florkowski:2017ruc, Florkowski:2017dyn, Florkowski:2018fap, Bhadury:2020cop, Bhadury:2020puc, Florkowski:2018ahw, Becattini:2009wh, Hattori:2019lfp, Weickgenannt:2020aaf,  Weickgenannt:2021cuo, Weickgenannt:2022qvh, Becattini:2018duy, Singh:2020rht, Florkowski:2019qdp, Shi:2020htn, Banerjee:2024xnd, Bhadury:2022ulr, Das:2022azr, Montenegro:2017lvf, Montenegro:2017rbu, Montenegro:2018bcf, Montenegro:2020paq, Biswas:2023qsw, Hu:2021lnx, Fukushima:2020ucl,Singh:2024cub, Sapna:2025yss}. For example, the classical description of spin is used to determine the structure of dissipative spin corrections~\cite{Bhadury:2020cop, Bhadury:2020puc}. On the other hand, the quantum-field theoretical aspect of spin hydrodynamics derived from the effective action principle describes causality and dissipation properties of polarized fluid~\cite{Montenegro:2017lvf, Montenegro:2017rbu, Montenegro:2018bcf, Montenegro:2020paq}. Similarly, other spin hydrodynamical approaches are based on various theoretical concepts such as thermodynamic equilibrium~\cite{Becattini:2009wh}, entropy-current analysis~\cite{Hattori:2019lfp, Biswas:2023qsw}, statistical operator~\cite{Hu:2021lnx}, non-local collisions~\cite{Weickgenannt:2020aaf, Weickgenannt:2021cuo, Weickgenannt:2022qvh}, chiral-kinetic theory~\cite{Shi:2020htn}, etc. Recently, a numerical method for spin hydrodynamics has been developed, and the longitudinal spin polarization is explored within the model~\cite{Singh:2024cub, Sapna:2025yss}. The spin sign puzzle has been attempted to solve by considering the gradient of chemical potential~\cite{Liu:2020dxg} and spin potential~\cite{Buzzegoli:2021wlg}. A chiral kinetic approach with AMPT initial conditions also predicts the same $P_z$ modulation as experimental data~\cite{Sun:2018bjl}. This approach considers the transverse component of vorticity and non-equilibrium effects due to spin degrees of freedom.\\

However, the observation of longitudinal spin polarization of $\Lambda$-hyperons is not limited only to its azimuthal angle dependence relative to the second-order event plane. The centrality, transverse momentum ($p_{\rm T}$), and rapidity dependence of longitudinal polarization are also observed at the RHIC and LHC energies~\cite{STAR:2019erd, Niida:2018hfw, ALICE:2021pzu, STAR:2023eck}. A simple boost-invariant blast wave (BW) hydrodynamic model with few freeze-out parameters explains the centrality and $p_{\rm T}$ dependence of longitudinal spin polarization measured at STAR Collaboration~\cite{STAR:2019erd}. It considers the kinematic vorticity associated with the velocity field without contribution from the temperature gradients and acceleration term. Furthermore, a hydrodynamic model MUSIC with AMPT initial conditions is used to estimate the centrality, $p_{\rm T}$ and rapidity dependence of longitudinal spin polarization of $\Lambda$-hyperons. This MUSIC+AMPT-based study considers both thermal shear and thermal vorticity into account. It predicts that the effect of thermal vorticity dominates over thermal shear when $\Lambda$ hyperon mass is used as the mass of the spin carrier at freeze-out to estimate the longitudinal spin polarization of $\Lambda$ and $\bar{\Lambda}$~\cite{ALICE:2021pzu}. In other cases, using the strange quark mass in place of the $\Lambda$ hyperon mass, the effect of thermal shear dominates over thermal vorticity. As a consequence, it predicts positive longitudinal spin polarization, which is similar to the observed experimental data. So, the MUSIC+AMPT-based study advocates that $\Lambda$ hyperons inherit the spin polarization of strange quarks. It is important to note that in both cases, the effect of hadronic interactions on $\Lambda$ hyperons spin polarization is not considered. Therefore, the particle spin polarization dynamics depend on the assumptions used in the model to calculate the polarization observable. So, it is crucial to validate the longitudinal spin polarization data with other hydrodynamic and transport models that consider the hadronic scattering into account.\\

In this study, we investigate the longitudinal spin polarization due to the thermal shear tensor in an ideal uncharged fluid, where temperature depends only on Bjorken time. In the earlier work~\cite{Becattini:2017gcx}, the authors have calculated the contribution of thermal vorticity on longitudinal 
spin polarization. Following Ref.~\cite{Becattini:2017gcx}, we derive an analytical expression relating the longitudinal mean 
spin vector to the elliptic flow due to the symmetric thermal shear tensor. To obtain the local $\Lambda$-hyperon 
spin polarization under the developed formulation, we have used hydrodynamic and 
multiphase transport models. The Eulerian Conservative High Order Quark Gluon Plasma (ECHO-QGP), and  Energy conservation Parallel scatting factOrization Saturation (EPOS4) models are used as a hydrodynamical-based approach and  AMPT model is used as a kinetic theory-based approach to estimate the longitudinal polarization of $\Lambda$ and $\bar{\Lambda}$-hyperons.  It is to be noted that the present study considers a longitudinal boost invariance scenario. Here, we investigate the azimuthal angle, centrality, and $p_{\rm T}$ dependence of longitudinal spin polarization of $\Lambda$-hyperons. \\

This paper is organized as follows. After a brief introduction in section~\ref{intro}, we discuss the formulation 
of elliptic flow-induced local spin polarization in section~\ref{formulation}. Section~\ref{model} briefly describes 
the event generators, ECHO-QGP, EPOS4, and AMPT. The results obtained from the model are discussed in  section~\ref{res}. Finally, the important findings are summarized in section~\ref{sum}.\\

\label{intro}
\section{Methodology}
\label{formulation}
The total mean spin vector of a spin-1/2 particle at local thermodynamic equilibrium is the sum of the contribution that arises due to the thermal vorticity and thermal shear~\cite{Becattini:2021suc, Becattini:2021iol, Palermo:2024tza, Becattini:2022zvf}.

\begin{equation}
S^{\mu}(p) \simeq  S^{\mu}_{\bar{\omega}}(p) + S^{\mu}_{\xi}(p)
\label{Stot}
\end{equation}

where, $S^{\mu}_{\bar{\omega}}(p)$ is the mean spin vector due to the thermal vorticity $\bar{\omega}$ and $ 
S^{\mu}_{\xi}(p)$ is the mean spin vector due to the thermal shear $\xi$. In this section, we calculate the analytical 
expression of longitudinal spin polarization for an ideal uncharged fluid due to thermal vorticity and thermal shear 
in subsection~\ref{vor} and ~\ref{shear}, respectively.

\subsection{Thermal vorticity}
\label{vor}
The mean spin vector of a spin-1/2 particle, characterized by its four-momentum $p$, is intricately linked to the thermal vorticity at leading order~\cite{Becattini:2013fla, Fang:2016vpj}, and is given by; 

\begin{equation}
S^{\mu}_{\bar{\omega}}(p) = -\frac{1}{8m}\epsilon^{\mu \rho \sigma \tau}p_{\tau} \frac{\int_{\Sigma} 
d\Sigma_{\lambda}p^{\lambda} \hspace{0.1cm} n_{F}(1-n_{F}) \bar{\omega}_{\rho \sigma}}{\int_{\Sigma} 
d\Sigma_{\lambda}p^{\lambda} \hspace{0.1cm}n_{F}},
\label{eq1}
\end{equation}

where $\bar{\omega}$ is the thermal vorticity, defined in terms of the anti-symmetric derivative of four 
temperature fields, given by; 

\begin{equation}
\bar{\omega}_{\mu \nu} = -\frac{1}{2}(\partial_{\mu}\beta_{\nu}-\partial_{\nu}\beta_{\mu})
\label{eq2}
\end{equation}

The relation between four temperature vector, $\beta^{\mu}$, and the four velocity $u^{\mu}$, along with the 
temperature  T, is expressed succinctly as $\beta^{\mu} = \frac{u^{\mu}}{T}$. Here, $n_{F} = 1/[\exp{(\beta. p - \sum_{j} {\mu_j q_j}/T)} + 1]$ is the Fermi-Dirac phase space distribution function.\\

Further, it is assumed that under the vicinity of the ideal uncharged fluid, the temperature vorticity,
\begin{equation}
\Omega_{\mu \nu} = \frac{1}{2}[\partial_{\mu} (Tu_{\nu})- \partial_{\nu} (Tu_{\mu})]
\label{Tvor}
\end{equation}
is conserved along the velocity i.e $\Omega_{\mu \nu}u^{\nu} = 0$ ~\cite{Becattini:2015ska, Becattini:2017gcx} and 
satisfies the relativistic Kelvin circulation theorem. As a consequence, if $\Omega_{\mu\nu}$ vanishes at initial time, 
it remains absent during the whole evolution of the ideal uncharged fluid. Thus by using this condition, we can 
obtain the thermal vorticity, $\bar{\omega}$, given as~\cite{Becattini:2015ska, Becattini:2017gcx}; 

\begin{equation}
\bar{\omega}_{\mu \nu} = \frac{1}{T}(A_{\mu} u_{\nu}- A_{\nu} u_{\mu})
\label{eq3}
\end{equation}
here, A is the four-acceleration field.\\

Now, using the Eq.~(\ref{eq3}) in Eq.~(\ref{eq1}), one can obtain the modified form of the spin vector for spin-1/2 particles, which is given as;

\begin{equation}
S^{\mu}_{\bar{\omega}}(p) = -\frac{1}{4m}\epsilon^{\mu \rho \sigma \tau}p_{\tau} \frac{\int_{\Sigma} 
d\Sigma_{\lambda}p^{\lambda} \hspace{0.1cm} A_{\rho}  \beta_{\sigma} n_{F}(1-n_{F}) }{\int_{\Sigma} 
d\Sigma_{\lambda}p^{\lambda} \hspace{0.1cm}n_{F}},
\label{eq4}
\end{equation}

For the ideal uncharged fluid, the equation of motion is defined as, 

\begin{equation}
A_{\rho} = \frac{1}{T} \nabla_{\rho} T = \frac{1}{T} (\partial_{\rho} - u_{\rho} u \cdot \partial) T .
\label{eq5}
\end{equation}

Replacing Eq.~(\ref{eq5}) in Eq. ~(\ref{eq4}), we get;

\begin{align}
S^{\mu}_{\bar{\omega}}(p) = -&\frac{1}{4mT}\epsilon^{\mu \rho \sigma \tau}p_{\tau} \bigg[ \frac{\int_{\Sigma} 
d\Sigma_{\lambda}p^{\lambda} \hspace{0.1cm}  \partial_{\rho}T \beta_{\sigma} n_{F}(1-n_{F}) }{\int_{\Sigma} 
d\Sigma_{\lambda}p^{\lambda} \hspace{0.1cm}n_{F}} \nonumber \\
&- \frac{\int_{\Sigma} d\Sigma_{\lambda}p^{\lambda} \hspace{0.1cm} \beta_{\sigma}u_{\rho} u.\partial T  n_{F}
(1-n_{F}) }{\int_{\Sigma} d\Sigma_{\lambda}p^{\lambda} \hspace{0.1cm}n_{F}} \bigg]
\label{eq6}
\end{align}

The inclusion of the term $\epsilon^{\mu \rho \sigma \tau} \beta_{\sigma}u_{\rho}$ in the second part of Eq.~(\ref{eq6}) results in this part becoming zero. 
Here, it must be noted that we assume the hypersurface $\Sigma$ is an isothermal decoupling hypersurface such that the temperature T is constant and can be taken out of the integral. This assumption is a very reasonable one at very high energy, where the chemical potentials basically vanish and the only thermodynamic parameter governing the hadronization transition is the temperature~\cite{Becattini:2021iol}. Therefore, the spin polarization tensor is 
solely determined by the first term.\\

Now, with the help of the relation 
$\frac{\partial}{\partial p^{\sigma}} n_{F} = - \beta_{\sigma} n_{F}(1-n_{F})$, 
Eq.~(\ref{eq6}) is rewritten as;

\begin{equation}
S^{\mu}_{\bar{\omega}}(p) = \frac{1}{4mT}\epsilon^{\mu \rho \sigma \tau}p_{\tau}  \frac{\int_{\Sigma} 
d\Sigma_{\lambda}p^{\lambda} \hspace{0.1cm}  \frac{\partial n_{F}}{\partial p^{\sigma}}  \partial_{\rho}T }
{\int_{\Sigma} d\Sigma_{\lambda}p^{\lambda} \hspace{0.1cm}n_{F}},
\label{eq7}
\end{equation}

By first solving the numerator of the above Eq.~\ref{eq7}, we get

\begin{align}
\int_{\Sigma} d\Sigma_{\lambda}p^{\lambda} \frac{\partial n_{F}}{\partial p^{\sigma}}  \partial_{\rho}T \; = 
\; &\frac{\partial}{\partial p^{\sigma}}\int_{\Sigma} d\Sigma_{\lambda}p^{\lambda}  n_{F}  \partial_{\rho}T 
\nonumber\\
&- \int_{\Sigma} d\Sigma_{\sigma} n_{F} \partial_{\rho}T.  \nonumber
\end{align}

At the relativistic limit, it is assumed that the decoupling hypersurface happens at $T = T_{c}$, where $T_{c}$ is 
the critical temperature. This entails that the normal vector to the hypersurface is the gradient of temperature, 
which forces the second integral of the above expression to vanish. Then, the final expression for the mean spin 
vector is given by, 

\begin{equation}
S^{\mu}_{\bar{\omega}}(p) = \frac{1}{4mT}\epsilon^{\mu \rho \sigma \tau}p_{\tau}   \frac{\frac{\partial}{\partial 
p^{\sigma}} \int_{\Sigma} d\Sigma_{\lambda}p^{\lambda} \hspace{0.1cm} n_F \; \partial_{\rho}T }{\int_{\Sigma} 
d\Sigma_{\lambda}p^{\lambda} \hspace{0.1cm}n_{F}}
\label{eq8}
\end{equation}

Here, we have considered an isochronous decoupling hypersurface, with a temperature field only dependent on the 
Bjorken time, $\tau$. Further, the decoupling hypersurface is described by the coordinates x, y, $\eta$, with  
$\tau$ as a constant.\\

Let us calculate the numerator of Eq.~(\ref{eq8}) putting $\rho = 0$, we obtain;

\begin{equation}
\int_{\Sigma} d\Sigma_{\lambda}p^{\lambda} \hspace{0.1cm} n_F \; \frac{d T}{d \tau} \cosh\eta.
\end{equation}

Re-evaluating Eq.~(\ref{eq8}), at $Y=0$, gives

\begin{equation}
S^{z}_{\bar{\omega}}(\textbf{p}_{T}, Y=0) \hat{\textbf{k}} \; \simeq \; - \frac{d T}{d \tau}\frac{1}{4mT} \hat{\textbf{k}} 
\frac{\partial}{\partial \phi} \log \int_{\Sigma} d\Sigma_{\lambda}p^{\lambda} n_{F}
\label{eq9}
\end{equation}

where $\phi$ is the azimuthal angle of the emitted particle in the transverse plane. Expanding the integral term of Eq.~(\ref{eq9}) 
in Fourier series of the azimuthal angular distribution at the freeze-out hypersurface, we obtain,
\begin{equation}
\int_{\Sigma} d\Sigma_{\lambda}p^{\lambda} n_{F} = \frac{dN}{2\pi E_{p}p_{\rm T} dp_{\rm T} } \bigg[ 1+ \sum_{n = 1}^{\infty} 2 v_{n} (p_{\rm T}) \cos n\phi  \bigg]
\label{FS}
\end{equation}

Substituting Eq.~\ref{FS} in Eq.~\ref{eq9} and keeping only the elliptic flow term, a simplified form of the longitudinal component of the mean spin vector can be obtained as;

\begin{align}
S^{z}_{\bar{\omega}}(\textbf{p}_{T}, Y=0)  \simeq - \frac{d T}{d \tau}\frac{1}{4mT} \frac{\partial}{\partial \phi} 2 v_{2}
(p_{\rm T}) \cos2\phi \nonumber  \\
= \frac{d T}{d \tau} \frac{1}{mT}  v_{2}(p_{\rm T}) \sin2\phi
\label{Sz}
\end{align}

In the rest frame of the particle, the longitudinal polarization vector $P^{*}_{z}$
can be calculated from the longitudinal mean spin vector $S^{*}_{z}$  by  using the relation,

\begin{equation}
P^{*}_{z} = 2S^{*}_{z}
\label{Pz}
\end{equation}

At the midrapidity, $S^{*}_{z}  = S_{z}$.\\

The Eq.~(\ref{Sz}) implies that the degree of longitudinal polarization depends on the temperature cooling rate, the 
second-order anisotropic flow coefficient, and the azimuthal angle of the emitted particles. In peripheral heavy-ion 
collisions, the initial spatial anisotropy of the overlap region is converted into an anisotropic azimuthal distribution 
in the momentum space of the final state particles. This anisotropy can be characterized in terms of Fourier 
coefficients, 

\begin{equation}
v_{2}  \equiv  < \cos[2(\phi - \Psi_{2})] >
\label{v2}
\end{equation}

where $\Psi_{2}$ is the azimuthal angle of the event plane for the second-order harmonics.\\

\subsection{Thermal shear}
\label{shear}

The mean spin vector of a spin-1/2 particle due to the thermal shear tensor is given by~\cite{Becattini:2021suc, Becattini:2021iol}; 

\begin{equation}
S^{\mu}_{\xi}(p) = -\frac{1}{4m}\epsilon^{\mu \rho \sigma \tau}p_{\tau} \frac{\int_{\Sigma} 
d\Sigma \cdot p  \; \;  n_{F}(1-n_{F}) \; \;  \hat{t}_{\rho} \frac{p^{\lambda}}{\varepsilon} \xi_{\lambda \sigma}}{\int_{\Sigma} 
d\Sigma \cdot p \hspace{0.1cm}n_{F}},
\label{Sp}
\end{equation}

where $\xi$ is the thermal shear, defined in terms of the symmetric derivative of four 
temperature fields, given by; 

\begin{equation}
\xi_{\mu \nu} = \frac{1}{2}(\partial_{\mu}\beta_{\nu} 
+ \partial_{\nu}\beta_{\mu})
\label{xi}
\end{equation}

here $\hat{t}_{\rho}$ is the direction of QGP in the center of mass frame; $ \varepsilon = \hat{t}_{\rho} \cdot p  = 
\sqrt{p^2 + m^2}$ is the energy, and  $n_F$ is the Fermi Dirac distribution function.\\

Following these, thermal shear tensor can be written as;

\begin{align}
\xi_{\mu \nu} = \frac{1}{2}(\partial_{\mu} \beta_{\nu} + \partial_{\nu}\beta_{\mu}) = \frac{1}{T} \left[\frac{1}{2} (\partial_{\mu} u_{\nu} + \partial_{\nu}u_{\mu}) \right] \nonumber \\  - \frac{1}{T^2} \left[\frac{1}{2}   (u_{\nu} \partial_{\mu}T + u_{\mu} \partial_{\nu}T) \right]
\label{xi2}
\end{align}
The symmetric shear tensor ($\Xi_{\mu \nu}$) can be decomposed as~\cite{Becattini:2021suc},

\begin{equation}
\Xi_{\mu \nu} = \frac{1}{2} (\partial_{\mu} u_{\nu} + \partial_{\nu}u_{\mu}) =  \frac{1}{2} (A_{\mu} u_{\nu} + A_{\nu} u_{\mu}) + \sigma_{\mu \nu} +\frac{1}{3} \theta \Delta_{\mu \nu}
\label{xi3}
\end{equation}
where A is the four acceleration field.\\

The shear stress tensor ($ \sigma_{\mu \nu}$) in Eq.~\ref{xi3} can be written as;
\begin{equation}
\sigma_{\mu \nu} = \frac{1}{2} (\nabla_{\mu} u_{\nu} + \nabla_{\nu}u_{\mu})  - \frac{1}{3} \theta \Delta_{\mu \nu}
\label{sigma}
\end{equation}
where $\theta$ = $\partial_{\mu} u^{\mu}$ is the expansion rate, 
$\Delta^{\mu \nu} = g^{\mu \nu} - u^{\mu} u^{\nu}$ is the transverse projector operator, and defining the orthogonal derivative $\nabla_{\mu} = \Delta_{\mu \nu} \partial^{\nu} = 
\partial_{\mu} - u_{\mu} u \cdot \partial$ \\

Further, the analytical expression for $\xi_{\mu \nu}$ from Eq.~\ref{xi2} is obtained as;
\begin{align}
\xi_{\mu \nu}  = \frac{1}{2T} \left[(A_{\mu} u_{\nu} + A_{\nu} u_{\mu}) + (\nabla_{\mu} u_{\nu} + \nabla_{\nu}u_{\mu})  \right] \nonumber \\ - \frac{1}{2T^2} \left[(u_{\nu} \partial_{\mu}T + u_{\mu} \partial_{\nu}T) \right]
\label{xi4}
\end{align}
Substituting the value of $\xi_{\lambda \sigma}$ in Eq.~\ref{Sp}, the mean spin vector is proportional to;
\begin{align}
S^{\mu}_{\xi}(p) \propto \epsilon^{\mu \rho \sigma \tau} \frac{\hat{t}_{\rho}}{\varepsilon} p_{\tau} p^{\lambda}  \bigg[ \frac{1}{2T} \left[(A_{\lambda} u_{\sigma} + A_{\sigma} u_{\lambda})  + (\nabla_{\lambda} u_{\sigma}   + \nabla_{\sigma}u_{\lambda}) \right] \nonumber \\ - \frac{1}{2T^2} \left[(u_{\sigma} \partial_{\lambda}T + u_{\lambda} \partial_{\sigma}T) \right] \bigg]
\label{sp2}
\end{align}
\begin{widetext}
Now, expanding the orthogonal derivative term in Eq.~\ref{sp2}, we get;
\begin{align}
S^{\mu}_{\xi}(p) \propto \epsilon^{\mu \rho \sigma \tau} \frac{\hat{t}_{\rho}}{\varepsilon} p_{\tau} p^{\lambda}  \bigg[  \frac{1}{2T} \left[ (A_{\lambda} u_{\sigma} + A_{\sigma} u_{\lambda}) + (\partial_{\lambda} u_{\sigma} - u_{\lambda} u \cdot \partial \; u_{\sigma} )  + ( \partial_{\sigma}u_{\lambda} -  u_{\sigma} u \cdot \partial \; u_{\lambda} ) \right] - \frac{1}{2T^2} (u_{\sigma} \partial_{\lambda}T + u_{\lambda} \partial_{\sigma}T)  \bigg]
\label{sp3}
\end{align}
Further simplification of the above equation gives, 
\begin{align}
S^{\mu}_{\xi}(p) \propto  \epsilon^{\mu \rho \sigma \tau}  \frac{\hat{t}_{\rho}}{\varepsilon} p_{\tau} \bigg[  \frac{1}{2T} \bigg[ (p \cdot A) u_{\sigma} + A_{\sigma} (p \cdot u)  + ( (p \cdot \partial) \; u_{\sigma} - (p \cdot u) (u \cdot \partial) \; u_{\sigma} ) +  (p^{\lambda} \partial_{\sigma}u_{\lambda} -  p^{\lambda} u_{\sigma} u \cdot \partial \; u_{\lambda} )  \bigg] - \nonumber \\ \frac{1}{2T^2} \left[( u_{\sigma} (p \cdot \partial) T + (p \cdot u) \partial_{\sigma}T) \right] \bigg]
\label{sp4}
\end{align}

Replacing acceleration four vector $A$ defined in Eq.~\ref{eq5}, Eq.~\ref{sp4} can be written as 

\begin{align}
S^{\mu}_{\xi}(p) \propto  \epsilon^{\mu \rho \sigma \tau}  \frac{\hat{t}_{\rho}}{\varepsilon} p_{\tau} \bigg[ \frac{1}{2T^2} \left[ (p^{\alpha} (\partial_{\alpha} - u_{\alpha} u \cdot \partial) T ) u_{\sigma} + (\partial_{\sigma} - u_{\sigma} u \cdot \partial)T \; \; (p\cdot u) \right] + \frac{1}{2T} \nonumber \\ \left[ ( (p \cdot \partial) \; u_{\sigma} - (p \cdot u) (u \cdot \partial) \; u_{\sigma} ) + (p^{\lambda} \partial_{\sigma}u_{\lambda} -  p^{\lambda} u_{\sigma} u \cdot \partial \; u_{\lambda} ) \right]  - \frac{1}{2T^2} \left[( u_{\sigma} (p \cdot \partial) T + (p \cdot u) \partial_{\sigma}T) \right] \bigg]
\label{mainS}
\end{align}

Using the relation between four momentum and four velocity $p^{\mu} = m u^{\mu}$ and $p \cdot u = m$, Eq.~\ref{sp3} can be transformed as 

\begin{align}
S^{\mu}_{\xi}(p) \propto  \epsilon^{\mu \rho \sigma \tau}  \frac{\hat{t}_{\rho}}{\varepsilon} p_{\tau} \left[ \frac{1}{2T^2} \bigg[ (p \cdot \partial T -  p \cdot \partial T ) u_{\sigma} + (\partial_{\sigma} - u_{\sigma} u \cdot \partial)T \; \; (p\cdot u) \right] + \frac{1}{2T} \nonumber \\ \left[ ( (p \cdot \partial) \; u_{\sigma} - (p  \cdot \partial) \; u_{\sigma} ) + (p^{\lambda} \partial_{\sigma}u_{\lambda} -  p^{\lambda} u_{\sigma} u \cdot \partial \; u_{\lambda} ) \right]  - \frac{1}{2T^2} \left[( u_{\sigma} (p \cdot \partial) T + (p \cdot u) \partial_{\sigma}T) \right] \bigg]
\label{mainS2}
\end{align}

\end{widetext}

With the help of the identity $u^{\lambda} \partial_{\sigma}u_{\lambda}$ = 0; Eq.~\ref{mainS2}, reduces to;

\begin{align}
S^{\mu}_{\xi}(p) \propto  \epsilon^{\mu \rho \sigma \tau}  \frac{\hat{t}_{\rho}}{2T^{2} \varepsilon } p_{\tau} \bigg[   (\partial_{\sigma} T - u_{\sigma} u \cdot \partial T) \; \; (p\cdot u)  \nonumber \\  -  \left[( u_{\sigma} (p \cdot \partial) T + (p \cdot u) \partial_{\sigma}T) \right] \bigg]
\label{mainS3}
\end{align}
Since we are considering the temperature field for Bjorken flow, the $\partial_{\sigma}T\longrightarrow 0$. On further simplification,  
Eq.~\ref{mainS3}, reduces to;

\begin{align}
    S^{\mu}_{\xi}(p) \propto  \epsilon^{\mu \rho \sigma \tau}  \frac{\hat{t}_{\rho}}{2T^{2} \varepsilon } p_{\tau} \bigg[  - u_{\sigma} u \cdot \partial T \; \; (p\cdot u)    -   u_{\sigma} (p \cdot \partial) T  \bigg]
        \label{mainS4}
\end{align}

On further simplification of $p \cdot u$ term, leads to;

\begin{align}
S^{\mu}_{\xi}(p) \propto - \epsilon^{\mu \rho \sigma \tau}  \frac{\hat{t}_{\rho}}{T \varepsilon } p_{\tau} (p \cdot \partial T ) \beta_{\sigma}
\label{mainS5}
\end{align}

Substituting Eq.~\ref{mainS5} in the Eq.~\ref{Sp}, the mean spin vector can be written as, 
\begin{equation}
S^{\mu}_{\xi}(p) = \frac{1}{4mT}\epsilon^{\mu \rho \sigma \tau}p_{\tau} \frac{\hat{t}_{\rho}}{\varepsilon}  \frac{\int_{\Sigma} 
d\Sigma \cdot p \;   (p \cdot \partial T) \;  n_{F}(1-n_{F}) \beta_{\sigma}}{\int_{\Sigma} 
d\Sigma \cdot p \hspace{0.1cm}n_{F}},
\label{first}
\end{equation}
Now, with the help of the relation 
$\frac{\partial}{\partial p^{\sigma}} n_{F} = - \beta_{\sigma} n_{F}(1-n_{F})$, 
Eq.~(\ref{first}) is rewritten as;

\begin{equation}
S^{\mu}_{\xi}(p) = \frac{-1}{4mT}\epsilon^{\mu \rho \sigma \tau}p_{\tau} \frac{\hat{t}_{\rho}}{\varepsilon}  \frac{\int_{\Sigma} 
d\Sigma \cdot p \;   (p \cdot \partial T) \; \frac{\partial}{\partial p^{\sigma}} n_{F}} {\int_{\Sigma} 
d\Sigma \cdot p \hspace{0.1cm}n_{F}},
\label{appeeq1}
\end{equation}

By solving the numerator of the above Eq.~\ref{appeeq1}, we get

\begin{align}
\int_{\Sigma} d\Sigma \cdot p \frac{\partial n_{F}}{\partial p^{\sigma}}  (p \cdot \partial T) \; =  \; \frac{\partial}{\partial p^{\sigma}}\int_{\Sigma} d\Sigma \cdot p  \;n_{F}  (p \cdot \partial T) \nonumber \\  -  \int_{\Sigma} d\Sigma \cdot p \;  n_{F} \;\partial_{\sigma} T - \int_{\Sigma} d\Sigma_{\sigma} n_{F} (p \cdot \partial T) 
\label{byparts1}
\end{align}
The second and third integral of the above equation vanishes with a similar argument as discussed above. Thus, Eq.~\ref{appeeq1}, 
becomes;
\begin{equation}
S^{\mu}_{\xi}(p) = \frac{-1}{4mT}\epsilon^{\mu \rho \sigma \tau}p_{\tau}  \frac{\hat{t}_{\rho}}{\varepsilon} \frac{ \frac{\partial}{\partial p^{\sigma}} \int_{\Sigma}  d\Sigma \cdot p \; n_F (p \cdot \partial T) \;}{\int_{\Sigma} d\Sigma \cdot p \hspace{0.1cm}n_{F}},
\label{appeeq2}
\end{equation}

At $\rho$ = 0 and $Y = 0$, Eq.~\ref{appeeq2} can be written as;

\begin{equation}
\int_{\Sigma} d\Sigma \cdot p \hspace{0.1cm} n_F \; \varepsilon \frac{d T}{d \tau} \cosh\eta.
\end{equation}
Hence, at , the $\rm cosh \eta  \approx 1$.
\begin{equation}
S^{\mu}_{\xi}(p) = \frac{-1}{4mT}\epsilon^{\mu \rho \sigma \tau}p_{\tau} \frac{ \frac{\partial}{\partial p^{\sigma}} \int_{\Sigma}  d\Sigma \cdot p \; n_F \frac{d T}{d \tau} \;}{\int_{\Sigma} d\Sigma \cdot p \hspace{0.1cm}n_{F}},
\label{appeeq3}
\end{equation}

Eq.~\ref{appeeq3}, follows the same form as Eq.~\ref{eq8}. Hence, the net contribution to the mean spin vector due to 
the thermal shear term is obtained as;

\begin{equation}
S^{z}_{\xi}(\textbf{p}_{T}, Y=0) \hat{\textbf{k}} \; \simeq \;  \frac{1}{4mT} \frac{d T}{d \tau} \hat{\textbf{k}} 
\frac{\partial}{\partial \phi} \log \int_{\Sigma} d\Sigma_{\lambda}p^{\lambda} n_{F}
\label{appeeq4}
\end{equation}
Following similar procedure as thermal vorticity and expanding the Eq.~(\ref{appeeq4}) in Fourier series in $\phi$ and keeping only the elliptic flow term, the longitudinal component of the mean spin vector can be expressed as;

\begin{align}
S^{z}_{\xi}(\textbf{p}_{T}, Y=0)  \simeq  \frac{1}{4mT} \frac{d T}{d \tau} \frac{\partial}{\partial \phi} 2 v_{2}
(p_{\rm T}) \cos2\phi \nonumber  \\
= \frac{-1}{mT} \frac{d T}{d \tau}   v_{2}(p_{\rm T}) \sin2\phi
\label{Sz2}
\end{align}

We found that the contribution of thermal shear on the mean spin vector is exactly equal but with an opposite sign for an 
ideal uncharged fluid, assuming the temperature evolution only with the Bjorken time.\\

In this study, the second-order anisotropic flow coefficient (elliptic flow) is 
obtained using ECHO-QGP, EPOS4, and the AMPT model. The hydrodynamic temperature evolution is obtained using the (3+1)D relativistic hydrodynamic model. In this work, the event-by-event elliptic flow coefficient is obtained using the event plane method ~\cite{Masera:2009zz}. A brief description of ECHO-QGP, EPOS4, and AMPT models and event generation using these models is provided in 
section~\ref{model}. \\

\section{Event Generation and DESCRIPTION OF THE MODEL}
\label{model}

Computational frameworks based on the modeling of theoretical physics with some approximation are commonly called 
``event generators''. These event generators play a crucial role in comprehending the physics of ultra-relativistic 
A$-$A, $p-$A, and $p-p$ collisions. At the fundamental level, these essential computational tools 
facilitate the connection between theoretical models and experimental data. As mentioned, the present study uses  ECHO-QGP, EPOS4, and AMPT event generators to estimate longitudinal polarization at ultra-relativistic heavy-ion collisions using elliptic flow and azimuthal angle of emitted particles. A brief introduction about these generators is provided in this section for completeness.

\subsection{ECHO-QGP}

ECHO-QGP is a relativistic hydrodynamic model used to study the dynamics of heavy-ion collisions. Here, we have used the setup of the ECHO-QGP with Israel-Stewart (IS) based (3+1)D viscous hydrodynamics model along 
Bjorken coordinates. In this study, we consider both shear and bulk viscosity to be finite, and these two viscosities are related to the squared speed of sound. The value of the shear viscosity to entropy density ratio 
is mentioned in Table~\ref{table1}. The initial conditions are obtained from the geometric Glauber model based on 
the energy density parameter. The parameters used to estimate the flow coefficients and, hence, the polarization 
in ECHO-QGP are discussed in Table~\ref {table1}. For the present study, we have taken the kinetic freeze-out temperature 
as 120 MeV for Au+Au collisions at $\sqrt{s_{\rm NN}}$ = 200 GeV and Pb+Pb collisions at $\sqrt{s_{\rm NN}}$ = 5.02 TeV as 
both of them are found to be nearly equal in the blast wave framework \cite{starBES} and our results don't change appreciably within a variation
of 20 MeV freeze-out temperature. The equation of state (EOS) of the system is a must to solve 
hydrodynamic equations. Although ECHO-QGP uses various EOS to describe the hydrodynamic evolution 
of the system, in our study, we used the realistic partial chemical equilibrium (PCE) EOS, which 
includes the comprehensive chemical composition of the hadronic species before the kinetic freeze-out. The freeze-out procedure is based on the  Cooper-Frye formalism. The ECHO-QGP succeeded in explaining the transverse momentum spectra and anisotropic flow coefficients (mostly the elliptic flow) in central and mid-
central collisions in the $p_{\rm T}$ range up to 1.5 - 2.0 GeV/c for the produced hadrons~\cite{Hirano:2010je, 
Chen:2011vt}. The detailed description and formalism implemented in the model can be found in 
Ref~\cite{DelZanna:2013eua}. The present study considers that hydrodynamic evolution occurs in a baryon-free 
region, it also neglects all possible electromagnetic effects and other possible fluctuations in the medium.

\begin{table}[htbp]
\centering
\renewcommand{\arraystretch}{1.3} 
\begin{tabular}{p{6cm} c c}
\hline \hline
\textbf{Parameters} & \textbf{Au+Au} & \textbf{Pb+Pb} \\
\hline \hline
Center of mass energy per nucleon, $\sqrt{s_{NN}}$ (GeV) & 200 & 5020 \\
Total inelastic scattering cross section, $\sigma_{NN}$ (mb) & 40 & 70 \\
Thermalization time, $\tau_{0}$ (fm/c) & 0.6 & 0.4 \\
Initial energy density, $e_0$ (GeV/fm$^3$) & 24.5 & 60.0 \\
Collision hardness, $\alpha$ & 1.0 & 1.0 \\
Shear viscosity to entropy density ratio, $\eta/s$ & 0.1 & 0.1 \\
Relaxation time coefficient for viscosity, $\tau_{\pi}$ (fm/c) & 3.0 & 3.0 \\
Shift of the rapidity plateau in pp collision, $\Delta_{\eta}$ & 3.0 & 3.0 \\
Width of the gaussian falloff in pp collision, $\sigma_{\eta}$ & 1.0 & 1.0 \\
Freeze-out temperature, $T_{\text{freeze}}$ (GeV) & 0.120 & 0.120 \\
\hline \hline
\end{tabular}
\caption{Parameter set used within the ECHO-QGP freeze-out routine.}
\label{table1}
\end{table}

\subsection{EPOS4}

The EPOS4 is an updated version of the Monte-Carlo event generator named EPOS~\cite{Werner:2013tya, Pierog:2013ria}. EPOS4 is developed with an array of unbiased parallel multiple-scattering formalisms. It incorporates the classical Gribov-Regge framework 
and the modern Dokshitzer-Gribov-Lipatov-Altarelli-Parisi (DGLAP) formalism. This amalgamation captures the 
parallel scattering within the S-matrix theory and characterizes the perturbative approach and saturation effects. 
Further, EPOS4 reproduces the perfect factorization and binary scaling in AA collisions by simultaneously 
considering the rigorous parallel scattering, energy-momentum sharing, AGK theorem, and factorization for hard 
processes by introducing saturation. In its description of hadronization, EPOS4 adopts a core-corona paradigm, 
where the high string density core hadronizes statistically while the low string density corona uses Lund string 
fragmentation~\cite{Pierog:2013ria}. The mechanism of the core-corona approach, which starts after some initial time 
$\tau_0$ = 0.4 
fm/c, has been discussed in detail in Ref.~\cite{Werner:2023jps}. The hot and dense QCD matter with 
low-$p_{\rm T}$ pre-hadrons, the ``core'', evolves with the (3+1)D viscous hydrodynamics using 
vHILL~\cite{Karpenko:2013wva}. It employs a straightforward crossover EOS calibrated with lattice QCD 
data~\cite{Werner:2013tya}. Based on the micro-canonical approach, as the system expands and cools down, it 
reaches the hadronization point at energy density $\epsilon_H \approx$ 0.57 GeV/fm$^{3}$. Meanwhile, pre-hadrons 
with high-$p_{\rm T}$ traverse through the core and form the ``corona''. However, these may 
still interact with the core hadrons through hadronic scatterings. The succeeding interactions between the 
formed hadrons are incorporated by employing the UrQMD as an after-burner. We have simulated a minimum bias of 
2 $\times 10^6$ events for Pb$+$Pb collisions at $\sqrt{s_{NN}}$ = 5.02 TeV, and 6 $\times 10^6$ events for Au$+$Au collisions at 
$\sqrt{s_{NN}}$ = 200 GeV using EPOS4. 

\subsection{AMPT}

A multi-phase transport model (AMPT) is a comprehensive approach that combines both initial partonic collisions 
and final hadronic interactions, effectively accounting for the transition between these distinct phases of 
matter. It is divided into four major components: first, the initialization of collisions is processed through the 
HIJING model ~\cite{Wang:1991hta}, parton transport is executed using the Zhang Parton Cascade (ZPC) model 
~\cite{Zhang:1997ej}, hadronization is based on the quark coalescence mechanism in string melting mode 
~\cite{Greco:2003xt}, and at last, hadron transport is conducted through a relativistic transport model (ART). 
Detailed descriptions of these models and their components are given in Ref.~\cite{Lin:2004en}. In the present 
study, we employ the string melting mode of AMPT for event generation as it effectively explains the anisotropic 
flow and particle spectra in the intermediate-$p_{\rm T}$ region, aligning with predictions provided by the quark 
coalescence mechanism for hadronization~\cite{Fries:2003vb, Fries:2003kq, Greco:2003mm}. To calculate the 
elliptic flow using AMPT, we adopted the averaged initial state from the Monte Carlo Glauber model, with 
parameters specified in Ref.~\cite{Loizides:2017ack}. The AMPT settings utilized are consistent with those 
reported in Refs.~\cite{Tripathy:2018bib} for heavy-ion collisions at LHC and RHIC energies. For our purpose, we 
generated 2 $\times 10^6$ minimum bias events for Pb$+$Pb collisions at $\sqrt{s_{NN}} = 5.02$ TeV, and 6 $\times 10^6$ events for Au$+$Au collisions at $\sqrt{s_{NN}} = 200$ GeV, employing identical settings within the AMPT framework.


\section{Results and Discussion}

\label{res} 
In this work, we estimate the longitudinal spin polarization vector ($P_{z}$) of $\Lambda$  ($\bar{\Lambda}$) hyperons as a function of azimuthal angle ($\phi$), collision centrality, and transverse momentum ($p_{\rm T}$). The longitudinal mean spin vector due to the thermal vorticity and thermal shear are given in Eq.~\ref{Sz} and Eq.~\ref{Sz2}, respectively. The longitudinal spin polarization depends on the decoupling hypersurface temperature, temperature gradient at hadronization, second-order anisotropic flow coefficient (i.e., the elliptic flow), mass, and the sine of the azimuthal angular distribution of emitted $\Lambda$  ($\bar{\Lambda}$) hyperons. We use a realistic (3+1)D viscous hydrodynamic ECHO-QGP, EPOS4, and a hybrid transport AMPT model to simulate heavy-ion collisions. Using these models, we obtain the elliptic flow and the azimuthal angle of emitted $\Lambda$  ($\bar{\Lambda}$) hyperons. The hydrodynamic temperature evolution at hadronization is obtained from the ECHO-QGP. In EPOS4, the elliptic flow coefficient is obtained using Eq~\ref{v2}. However, in AMPT and ECHO-QGP, there is a provision for making the reaction plane angle $\Psi_{2}$ = 0, although it is nontrivial in experiments. Thus, one can obtain the elliptic flow coefficients as $v_{2}  =  < \cos2\phi >$, here, the average is taken over all $\Lambda$ produced in an event. The choice of centrality selection in the ECHO-QGP  model is obtained for fixed impact parameter values based on Monte-Carlo Glauber model estimation~\cite{Loizides:2017ack}. However, for EPOS4 and AMPT models, the centrality selection is performed with the geometrical slicing of the impact parameter distribution.\\

\begin{figure}[ht!]
\includegraphics[scale=0.45]{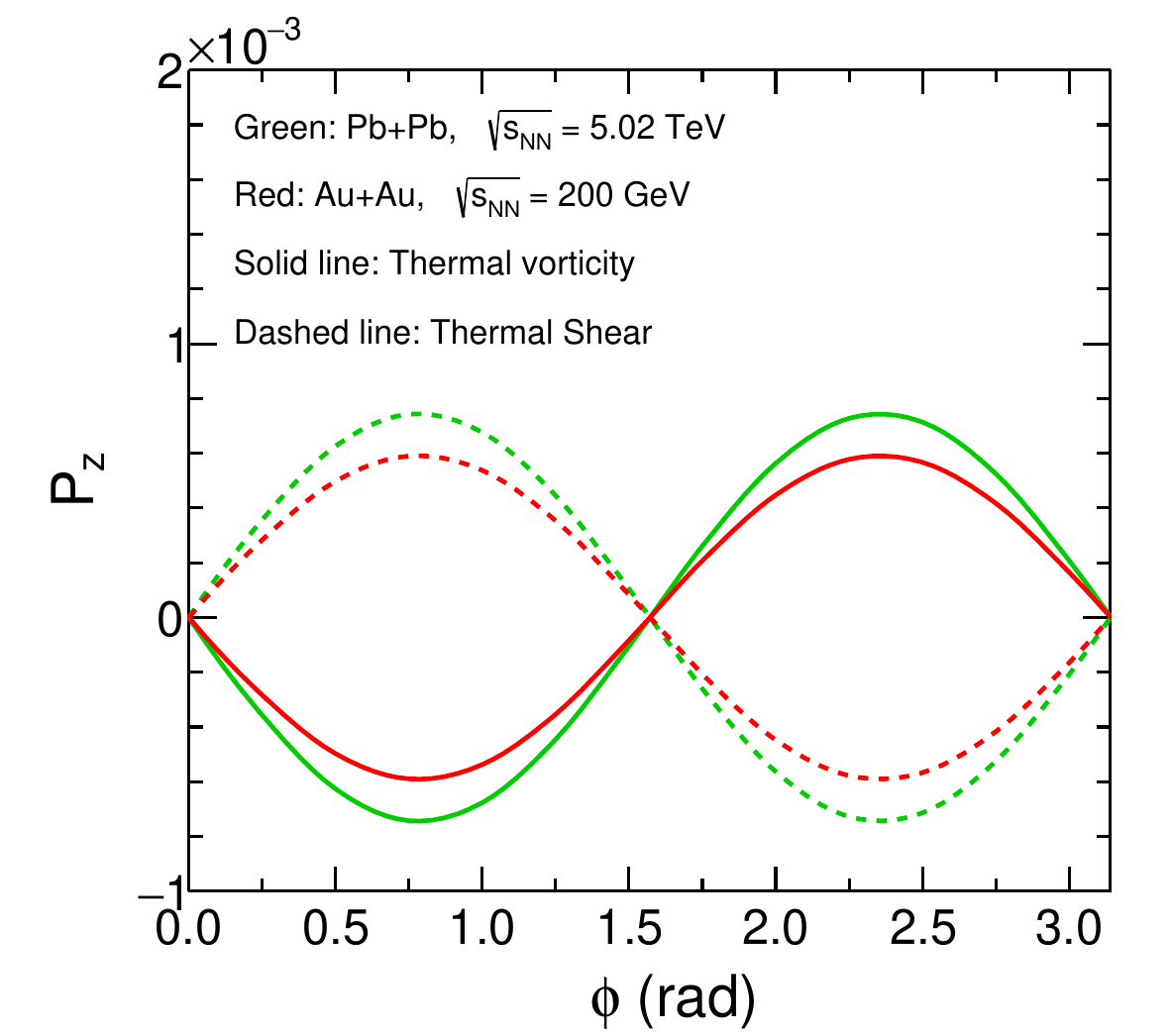}            
\caption{(Color online) The longitudinal component of $\Lambda$ and $\bar{\Lambda}$ spin polarization ($P_{z}$) as a function of azimuthal angle ($\phi$) obtained from ECHO-QGP model simulation for $\sqrt{s_{NN}}$ = 5.02 TeV and  $\sqrt{s_{NN}}$ = 200 GeV in Pb$+$Pb and Au$+$Au collisions, respectively.} 
\label{fig:pzphi}
\end{figure}

\begin{figure}[ht!]
\includegraphics[scale=0.45]{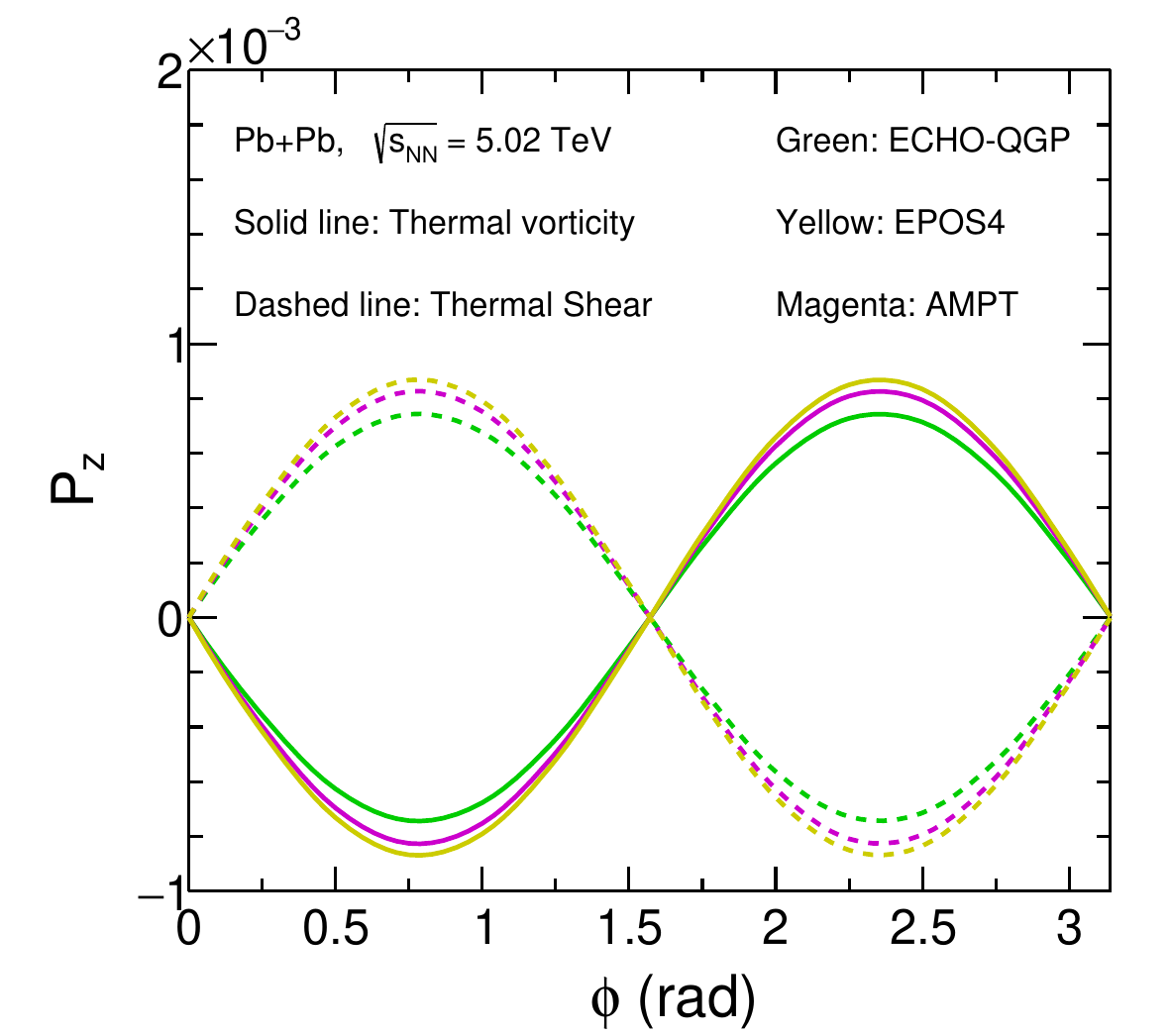}          
\caption{(Color online) The longitudinal component of $\Lambda$ and $\bar{\Lambda}$ spin polarization ($P_{z}$) as a function of azimuthal angle ($\phi$) obtained from ECHO-QGP, EPOS4, AMPT model simulation for $\sqrt{s_{NN}}$ = 5.02 TeV in Pb$+$Pb collisions.} 
\label{fig:pzphimodel}
\end{figure}

Figure~\ref{fig:pzphi} shows $P_{z}$ as a function of $\phi$ for $\Lambda$-hyperons obtained using the ECHO-QGP 
hydrodynamic model simulation. The green solid and dashed line shows  $P_{z}$ due to the thermal vorticity and the 
thermal shear, respectively, in Pb+Pb collisions at $\sqrt{s_{NN}}$ = 5.02 TeV for (30-50)\% with $p_{\rm T}$ range 0.15 
$ < p_{\rm T} <$ 3.0 GeV/c. Similarly, the red solid and dashed line shows the $P_{z}$ due to the thermal vorticity and 
the thermal shear, respectively, for Au+Au collisions at  $\sqrt{s_{NN}}$ = 200 GeV with the same $p_{\rm T}$ and 
centrality classes. Figure~\ref{fig:pzphi} indicates the sine modulation of $P_{z}$ with $\phi$. The value of $P_{z}$ 
due to the thermal shear and thermal vorticity are found to be exactly equal with an opposite sign. A similar 
observation is reported in Refs.~\cite{Yi:2021ryh, Florkowski:2021xvy}. It is evident that considering the thermal shear 
in place of thermal vorticity qualitatively reproduces the correct sign of sine modulation of longitudinal polarization 
as observed at the STAR experiment~\cite{STAR:2019erd}.  Similarly, Fig.~\ref{fig:pzphimodel} shows $P_{z}$ as a function of 
$\phi$ for three different models used in the study for Pb$+$Pb collisions at (30-50)\% centrality class and $p_{\rm T}$ range 0.15 $ < p_{\rm T} <$ 3.0 GeV/c. Figure~\ref{fig:pzphi} investigates the $\sqrt{s_{NN}}$ dependence of the longitudinal polarization using the ECHO-QGP hydrodynamic model. On the other hand, Fig.~\ref{fig:pzphimodel} aims to predict the model (hydrodynamic vs. transport) dependence of the longitudinal spin polarization at a given $\sqrt{s_{NN}}$. It explores how different models with different physics assumptions and implementations predict the same observable at the same collision energy.  A very marginal difference in $P_{z}$ as a function of $\phi$ is found between the hydrodynamic and transport models. This dual approach strengthens the investigations regarding the robustness and sensitivity of longitudinal polarization. \\

It is noteworthy to mention that the contribution of thermal shear and thermal vorticity to the longitudinal mean spin 
vector are exactly equal in magnitude, but an opposite sign is obtained only using the equation of motion for an ideal 
uncharged fluid. This simple solution of relating the longitudinal mean spin vector with elliptic flow is obtained by 
assuming the temperature evolution only with the Bjorken time. However, considering the temperature gradient, chemical 
potential gradient, and electromagnetic field terms in the mean spin vector, a finite longitudinal spin polarization is 
expected~\cite{Yi:2021ryh,Wu:2022mkr,Yi:2023tgg}.\\ 

\begin{figure}[ht!]
\includegraphics[scale=0.45]{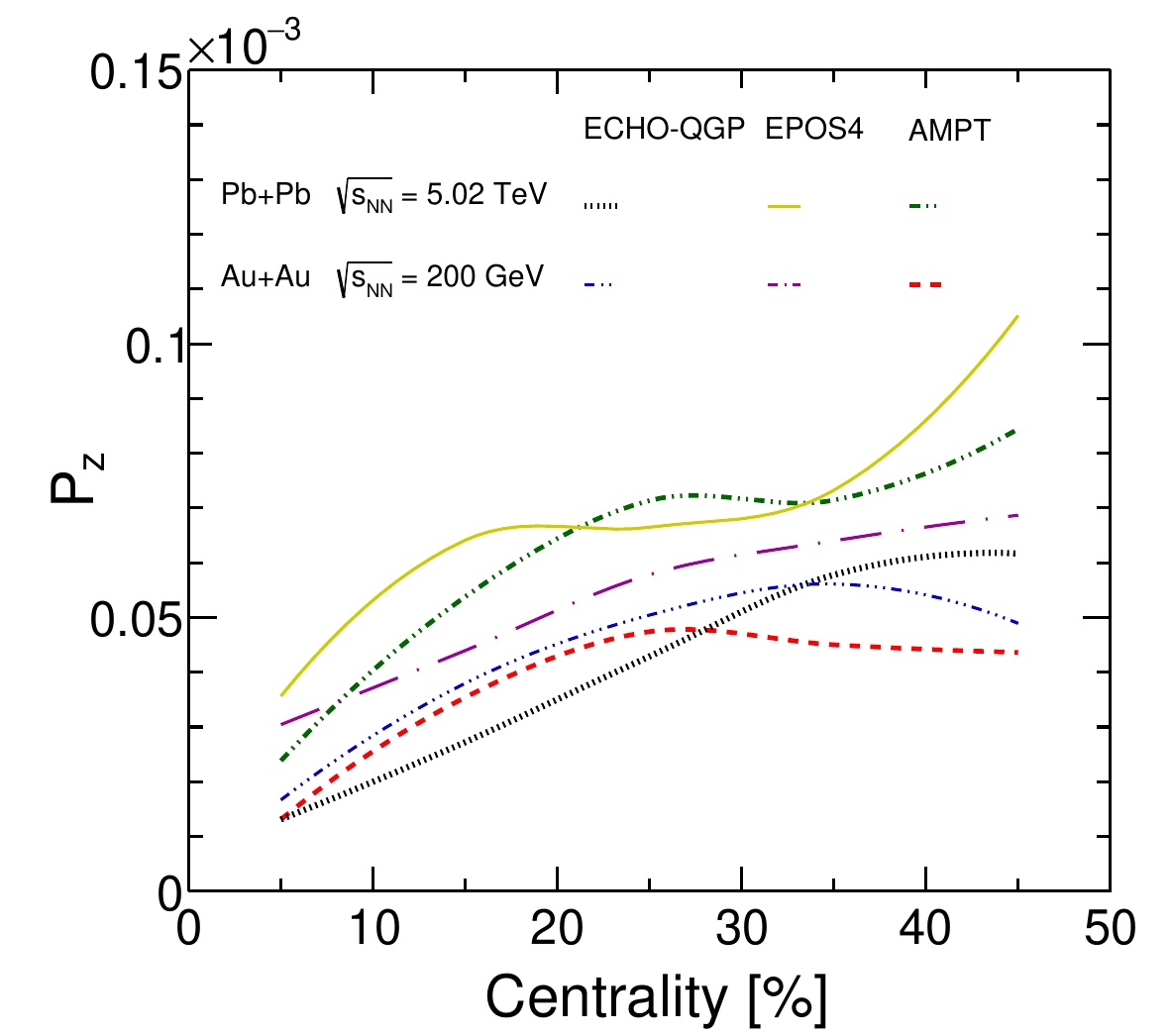}            
\caption{(Color online) The longitudinal component of $\Lambda$ and $\bar{\Lambda}$ polar ization ($P_{z}$) as a function of collision centrality obtained from ECHO-QGP, EPOS4, and AMPT model simulation for $\sqrt{s_{NN}}$ = 5.02 TeV and  $\sqrt{s_{NN}}$ = 200 GeV in Pb$+$Pb and Au$+$Au collisions, respectively.} 
\label{fig:1}
\end{figure}

The centrality dependence of $P_{z}$ for $\Lambda$-hyperons obtained from hydrodynamic and transport models is shown in 
Fig.~\ref{fig:1} for $p_{\rm T}$ range 0.15 $ < p_{\rm T} <$ 3.0 GeV/c in Au$+$Au collisions at $\sqrt{s_{NN}}$ = 200 
GeV and Pb$+$Pb collisions at  $\sqrt{s_{NN}}$ = 5.02 TeV, taking only thermal vorticity contribution. The positive $P_{z}$ shown in Fig.~\ref{fig:1} is the 
manifestation of two negative signs appearing in Eq.(11); one is from the temperature cooling rate, and the other one is 
from sin2$\phi$.  The term sin2$\phi$ in Eq.~\ref{Sz} is calculated by taking the sine of the average over all azimuthal angles carried by the $\Lambda$ hyperons produced in an event. It is observed that the magnitude 
of longitudinal spin polarization increases from most-central to mid-central collisions. The observed increase in signal 
with decreasing centrality can be attributed to the rising contributions of elliptic flow in peripheral collisions. In 
most central collisions, the polarization seems to disappear, whereas in such collisions, the elliptic flow might be 
significant due to the initial density fluctuations.

ECHO-QGP predicts a slightly higher value of $P_z$  in top (0-30)\% central  Au$+$Au collisions at $\sqrt{s_{NN}}$ = 
200 GeV compared to the Pb$+$Pb collision at $\sqrt{s_{NN}}$ = 5.02 TeV. Furthermore, around (30-50)\% mid-central 
collisions, the degree of longitudinal polarization obtained from the ECHO-QGP model dominates for Pb$+$Pb collisions 
compared to the Au$+$Au collisions. The AMPT model predicts a slightly higher value of $P_z$ for Pb$+$Pb, 
$\sqrt{s_{NN}}$ = 5.02 TeV as compared with the Au$+$Au, $\sqrt{s_{NN}}$ = 200 GeV for all centrality classes. The trend 
of longitudinal polarization with centrality for EPOS4 is the same as AMPT, though both are based on different 
formulations. The current formulation integrated with AMPT and EPSO4 predicts change in the magnitude of 
polarization, $P_z$ for $\Lambda$ and $\bar{\Lambda}$, with  center-of-mass energy ($\sqrt{s_{NN}}$).  
However, ECHO-QGP based results for the $P_z$ of $\Lambda$ and $\bar{\Lambda}$ are inconsistent with $\sqrt{s_{NN}}$. 
The small variations between the hydrodynamic and transport models arise due to different evolution and 
hadronization processes are considered in these models. Here, we have observed that the magnitude of the collision 
center of the mass-energy dependence of elliptic flow-induced polarization along the beam direction is weak as compared 
to the global spin polarization reported in Ref.~\cite{STAR:2017ckg}.\\

\begin{figure}[!ht]
\includegraphics[scale=0.45]{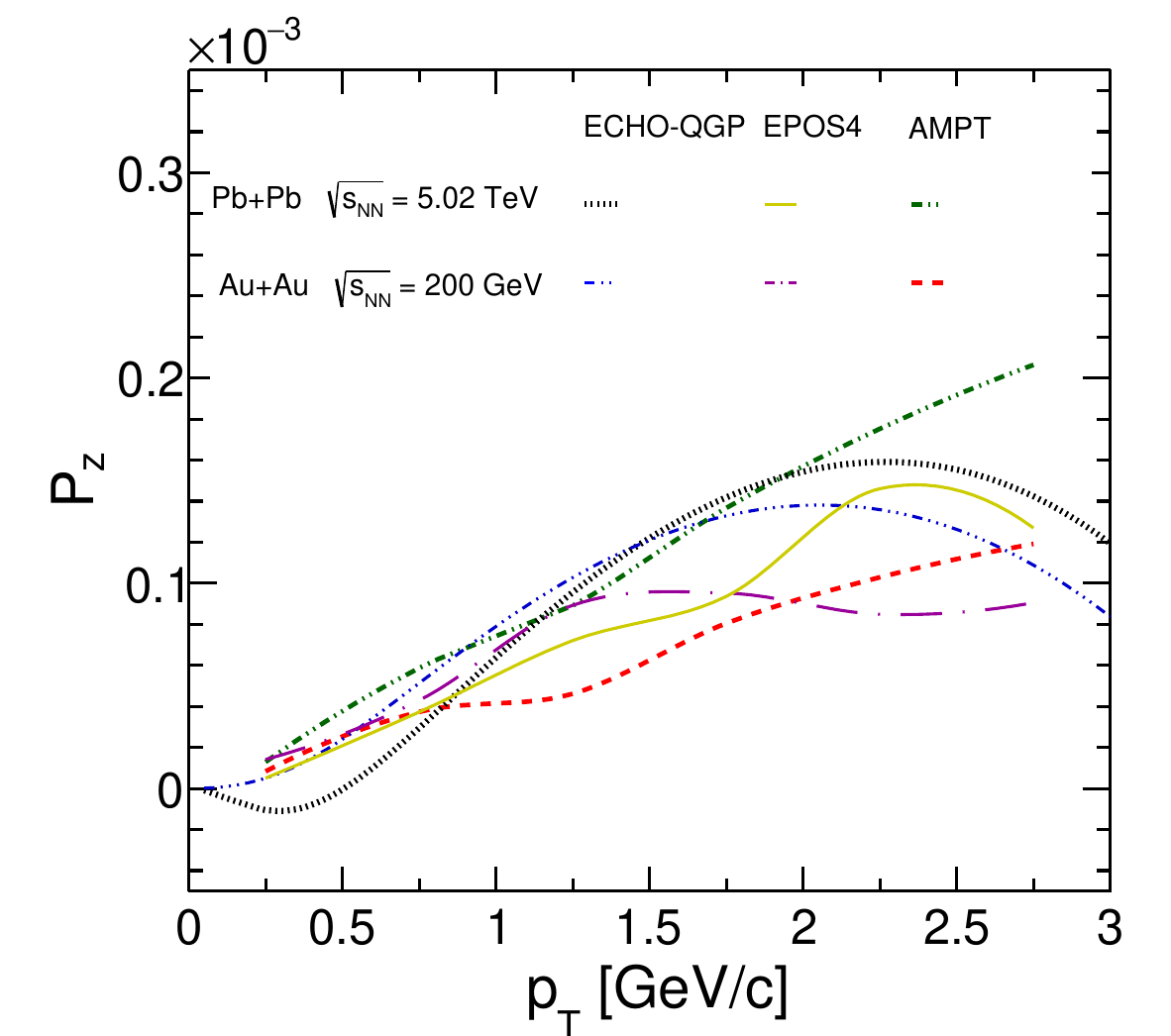}      
\caption{(Color online) The longitudinal component of $\Lambda$ and $\bar{\Lambda}$ polarization ($P_{z}$) as a function of transverse momentum ($p_{\rm T}$) obtained from ECHO-QGP, EPOS4, and AMPT model simulation for $\sqrt{s_{NN}}$ = 5.02 TeV and  $\sqrt{s_{NN}}$ = 200 GeV in Pb$+$Pb and Au$+$Au collisions, respectively for (30-50)\% centrality classes. }
\label{fig:2}
\end{figure}

Since the elliptic flow depends on the centrality and $p_{\rm T}$, the spin polarization along the beam direction also 
exhibits the $p_{\rm T}$ dependence. Figure \ref{fig:2} indicates a $p_{\rm T}$ dependence of the 
elliptic flow-induced polarization for $\Lambda$ ($\bar{\Lambda}$) hyperons in (30-50)\% centrality bin by taking only the
thermal vorticity contribution.  It is observed 
that the longitudinal spin polarization increases for $p_{\rm T} \lesssim $ 2.0 GeV/c and decreases towards $p_{\rm T} 
\gtrsim $ 2.0 GeV/c for ECHO-QGP. Further, the ECHO-QGP predicts the negative longitudinal polarization at low $p_{\rm 
T}$ for Pb+Pb collisions. Because the elliptic flow obtained from ECHO-QGP has negative values for $\Lambda$ 
($\bar{\Lambda}$) hyperons at low$-p_{\rm T}$. While for AMPT, $P_z$ increases with $p_{\rm T}$, and this increasing 
trend of beam-induced polarization as a function of $p_{\rm T}$ is due to the elliptic flow behavior obtained from the 
AMPT. For EPOS4, $P_z$  exhibits a non-monotonic behavior for $\Lambda$ ($\bar{\Lambda}$) hyperons with  $p_{\rm T}$.

\section{Summary}
\label{sum}

In summary, we estimate the longitudinal component of spin polarization vector for $\Lambda$ and $\bar{\Lambda}$ hyperons in Au$+$Au collisions at $\sqrt{s_{NN}}$ = 200 GeV and Pb$+$Pb collisions at  $\sqrt{s_{NN}}$ = 5.02 TeV. We use a (3+1)D relativistic viscous hydrodynamic model, such as ECHO-QGP, EPOS4, and a hybrid transport AMPT model, to investigate the spin polarization along the beam direction. We obtain the longitudinal spin polarization from the elliptic flow coefficients in a longitudinal boost scenario with no initial state fluctuations. We assume the spin degrees of freedom are in local thermodynamic equilibrium with the ideal fluid. The results show a centrality dependence of longitudinal spin polarization with increasing magnitude from most-central to mid-central collisions, which tends to decrease towards peripheral collisions. We observe a transverse momentum dependence of longitudinal spin polarization in both systems at relativistic energies. It is observed that the longitudinal spin polarization increases with  $p_{\rm T} \lesssim $ 2.0 GeV/c and seems to decrease towards $p_{\rm T} \gtrsim $ 2.0 GeV/c. 
It is worth mentioning that the spin polarization is primarily governed by the thermal vorticity field and/or thermal shear tensor. The evolution of the thermal vorticity field and thermal shear depends on the chosen initial thermodynamic conditions. Therefore, the present study is sensitive to the chosen initial conditions for the hydrodynamic and transport models simulation and the temperature gradient at the decoupling stage.
It is to be noted that the model parameters and initial conditions considered in our calculation are commonly used in various studies.\\

The longitudinal spin polarization obtained in this study is due to the second-order anisotropic flow coefficients, showing the quadrupole structure of the vorticity field in the transverse plane. However, the higher-order anisotropic flow harmonics also have similar vorticity structures in the transverse plane and contribute to the longitudinal spin polarization. Recently, the STAR Collaboration reported such an observation in Ref.~\cite{STAR:2023eck}. The sine modulation of $P_{z}$ relative to the third-order event plane exhibits a sextupole pattern of vorticity induced by triangular flow in isobar collisions. It is important to note that apart from thermal vorticity and thermal shear contribution, the acceleration term (fluid acceleration minus the temperature gradient), the gradient of axial chemical potential over temperature (so-called spin hall effect), and the external electromagnetic field will also contribute to the mean spin vector~\cite{Yi:2021ryh, Wu:2022mkr, Yi:2023tgg}. However, the contribution of the acceleration term and axial chemical potential to the longitudinal spin polarization is found to be negligible at a higher center of mass energy obtained through hydrodynamic simulations~\cite{Wu:2022mkr}. Since the electromagnetic field decays rapidly in relativistic heavy-ion collisions, its contribution to the local spin polarization is also assumed to be negligible~\cite{Yi:2021ryh, Wu:2022mkr}. However, the axial chemical potential mainly contributes to the helicity polarization and could be implemented to probe the local-parity violation in QCD matter at finite temperature~\cite{Yi:2021ryh, Becattini:2020xbh}. Apart from these, there could be various other factors that affect the local spin polarization of $\Lambda$ hyperons.  Such as the feed-down of higher excited states and considering particle spin as an independent degree of freedom in each stage of the relativistic spin magneto-hydrodynamic framework may provide a comprehensive understanding of the local spin polarization.\\

More insightful investigations are required to comprehensively understand the global and/or local quark spin 
polarization in the QGP medium. These studies may comprise the spin-transfer mechanism at the hadronization stage, the 
relaxation time required to convert the vorticity to particle polarization, the effect of hadronic rescattering on the 
spin polarization, etc. With the current Run 3 high statistics data samples of ALICE, more differential and precision 
measurements can be performed on the local and global spin polarization of single and multi-strange baryons. With the precise 
vertexing and low $p_{T}$ measurement capabilities for heavy flavor hadrons in Run 3 of ALICE, it would be interesting 
to investigate the global and local spin polarization dynamics of heavy flavor hadrons. This may reveal some of the open 
questions, such as: (i) Are the heavy flavor hadrons thermalized with the medium? (ii) Is the spin polarization achieved at 
the quark-gluon plasma phase or the hadronic phase? (iii) Does the hadron spin polarization depend upon the quantum number of 
the constituent quark species?\\

\section*{Acknowledgement}
The authors would like to acknowledge some fruitful discussions with Ronald Scaria and Suraj Prasad during the preparation of the manuscript. Bhagyarathi Sahoo acknowledges the financial aid from CSIR, Government of India. The authors gratefully acknowledge the DAE-DST, Government of India, funding under the mega-science project "Indian Participation in the ALICE experiment at CERN" bearing Project No. SR/MF/PS-02/2021-IITI (E-37123).

\appendix

\end{document}